\documentclass[showpacs,twocolumn,prb]{revtex4}
\usepackage{makeidx}
\usepackage{amsfonts}
\usepackage[latin9]{inputenc}
\usepackage{amsmath}
\usepackage{amssymb}
\usepackage{graphicx}

\begin{document}

\title{Magnetic field at the center of a vortex: a new criterion for the
classification of the superconductors}
\author{Jun-Ping Wang}
\affiliation{Department of physics, Yantai university, Yantai 264005, People's Republic of China}

\begin{abstract}
Magnetic response of a superconductor depends on the thermodynamic stability
of vortex in the material. Here we show that the vortex stability has a
close relation with the ratio of the magnetic field at the vortex core
center to the thermodynamic critical field. This finding provides a new
criterion for the classification of the superconductors according to their
magnetic responses.
\end{abstract}

\maketitle

Type-1 and type-2 superconductors exhibit different magnetic responses to
externally applied magnetic field. Whether there exists stable vortex is the
distinction between two types of superconductors \cite{Abrikosov}. Due to the formation of
stable vortices, mix state appears in a type-2 superconductor under a
certain range of external field. As a consequence of thermodynamic
instability of vortex, only macroscopic Meissner state and fully normal
state or the state of the coexistence of these two states (intermediate
state) can exist in a type-1 material.

To date, the identifications of types of superconductors are based on the
following three considerations: (1) to calculate the surface energy of a
superconductor under the thermodynamic critical magnetic field \cite{GL origin};
(2) to compare the lower (higher) critical field at which the vortex entry into
(exit from) the superconductor to the thermodynamic critical magnetic field \cite{Landaubook,kogan};
(3) to analyze the interaction between vortices \cite{kramer,jacobs,brandt,speight,mohamed,
alford,chaves}. All these three considerations
show that, the magnetic response of a superconductor is determined by a
dimensionless parameter $\kappa $, which is defined as the ratio of the
penetration depth to the coherence length. And the critical value of the
Ginzburg-Landau (GL) parameter $\kappa _{c}=1/\sqrt{2}$ represents a
boundary between two types of superconductors.

In the present work, we revisit the problem of the thermodynamic stability
of the vortex in a superconductor. We show that, there is a simple and
rigorous relation between the magnetic field at the center of a vortex and
the thermodynamic stability of vortex. Concretely, when the magnetic field
at the center of vortex is larger than the thermodynamic critical field of
the material, vortex is unstable and the superconductor is of type-1. On the
contrary, when the magnetic field at the vortex core center is smaller than
the thermodynamic field, vortex is stable and the superconductor is of
type-2. The critical value, while the field at the core center equals to the
thermodynamic critical field, represents a boundary between two cases. Since
the magnetic response of a superconductor is completely determined by the
the stability of vortex, our finding can serve as a new criterion for the
classification of superconductors.

Vortex solution can always be constructed in the phenomenological
Ginzburg-Landau model for superconductivity. Here we use the GL free energy \cite{type2book}%
\begin{equation*}
f=f_{n0}+\frac{\hbar ^{2}}{2m^{\ast }}\left( \nabla \left\vert \Psi
\right\vert \right) ^{2}+\frac{m^{\ast }c^{2}}{32\pi ^{2}e^{\ast
2}\left\vert \Psi \right\vert ^{2}}\left( \nabla \times \mathbf{B}\right)
^{2}
\end{equation*}%
\begin{equation}
+V(\left\vert \Psi \right\vert ^{2})+\frac{\mathbf{B}^{2}}{8\pi },
\label{energy}
\end{equation}%
where $f_{n0}$ is the free energy density of the body in the normal state in
the absence of the external field, $e^{\ast }$ and $m^{\ast }$ are the
effective mass and charge of the Cooper pair, $\left\vert \Psi \right\vert $
is the modulus of the superconducting order parameter, $\mathbf{B}$ is the
magnetic field, $V(\left\vert \Psi \right\vert ^{2})=\alpha \left\vert \Psi
\right\vert ^{2}+\beta /2\left\vert \Psi \right\vert ^{4}$. The free energy $%
\left( \ref{energy}\right) $ is equivalent to the usual GL model in which
the gauge potential $\mathbf{A}$ and the order parameter $\Psi $ are the
functions to describe the superconductivity, and $\mathbf{B}=\nabla \times
\mathbf{A}$. We consider an isolate vortex in an infinite sample, and use
instead of the variable $r$, the function $\left\vert \Psi \right\vert $ and
$\mathbf{B}$ the dimensionless quantities $\rho =r/\lambda ,\ \left\vert
\psi \right\vert =\left\vert \Psi \right\vert /\Psi _{0},\ B=\left\vert
\mathbf{B}\right\vert \mathbf{/}H_{c}$\textbf{, }where $\lambda =\left(
m^{\ast }c^{2}\beta /4\pi e^{\ast 2}\left\vert \alpha \right\vert \right)
^{1/2}$ is the penetration depth$,\ \Psi _{0}=\left( -\alpha /\beta \right)
^{1/2},\ H_{c}=\left( 4\pi \alpha ^{2}/\beta \right) ^{1/2}$ is the
thermodynamic critical magnetic field. Radially symmetric vortex solution
can be found by solving the following equations%
\begin{equation*}
-\frac{1}{\kappa ^{2}}\frac{1}{\rho }\frac{d}{d\rho }\left( \rho \frac{%
d\left\vert \psi \right\vert }{d\rho }\right) +\frac{1}{2\left\vert \psi
\right\vert ^{3}}\left( \frac{dB}{d\rho }\right) ^{2}-\left\vert \psi
\right\vert +\left\vert \psi \right\vert ^{3}=0,
\end{equation*}%
\begin{equation}
\frac{1}{\rho }\frac{d}{d\rho }\frac{\rho }{\left\vert \psi \right\vert ^{2}}%
\frac{dB}{d\rho }=B,  \label{vortex}
\end{equation}%
where $\kappa =\lambda /\xi $ is the GL parameter, $\xi =\hbar /\sqrt{%
2m^{\ast }\left\vert \alpha \right\vert }$ is the coherence length. Far from
the vortex core, the order parameter approaches the ground state value and
the magnetic field vanishes to guarantee the Meissner state,%
\begin{equation}
\left\vert \psi (\infty )\right\vert =1,\ B\left( \infty \right) =0.
\label{infinite}
\end{equation}

\begin{figure}[tbp]
\begin{center}
\includegraphics[width=3.35in,keepaspectratio]{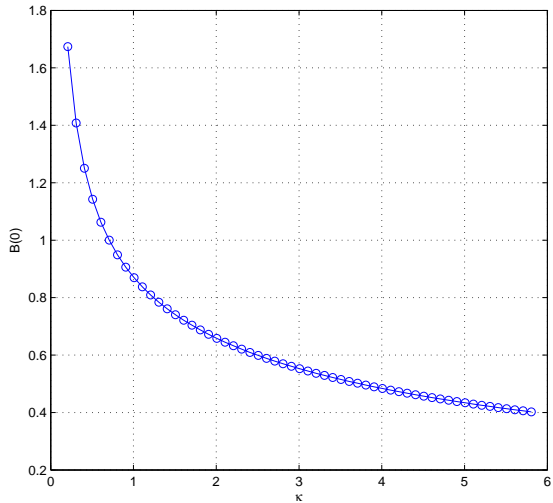}
\end{center}
\caption{(Color online) Magnetic field at the center of a vortex $B(0)$ as
function of the Ginzburg-Landau parameter $\protect\kappa$. $B(0)=1$
corresponds to $\protect\kappa=1/\protect\sqrt{2}$. }
\label{fig1}
\end{figure}

The fact that the vortex carries a nontrivial topological charge leads to
the important conclusion that the flux carried by the vortex is quantized in
unit of $hc/e^{\ast }$. Here we consider an isolated vortex line
enclosing a single flux quantum, which is expected to have the lowest free
energy. In this case, asymptotic behaviors of $\left\vert \psi \right\vert $
and $B$ near the center of the vortex can be deduced from the quantization
of flux $\int B\rho d\rho =\sqrt{2}/\kappa $ and Eqs. (\ref{vortex}),%
\begin{eqnarray}
\left\vert \psi \right\vert &=&a\rho +...,\   \label{core} \\
B &=&B(0)-\frac{a^{2}}{\sqrt{2}\kappa }\rho ^{2}+...,  \notag
\end{eqnarray}%
where $a$ is the slope of $\left\vert \psi \right\vert $, $B(0)$ is the
magnetic field at the center of the vortex. With boundary conditions of $%
\left\vert \psi \right\vert $ and $B$ (\ref{infinite}) and (\ref{core}),
Eqs. (\ref{vortex}) can be solved. Then the vortex solution and $a$, $B(0)$
are determined \cite{code}. Our results are more precise and comprehensive than
those of previous work \cite{tholfsen,khanra}. In addition, precise solutions of
GL vortex can also be obtained by using the iteration method
developed by Brandt \cite{brandt2}.

We note that the difference of vortex solutions originates from the
difference of the GL parameter $\kappa $. This means that $B(0)$, the
magnetic field at the center of vortex, should be a function of the GL
parameter $\kappa $,
\begin{equation}
B\left( 0\right) =B(0)(\kappa ).
\end{equation}%
Further, considering the fact that the flux quantization is independent of
the GL parameter $\kappa $, we then conclude that the magnetic field at the
vortex center, $B\left( 0\right) $, $must$ be a monotonic function of the GL
parameter $\kappa $.

In figure 1, magnetic field at the center of a vortex as a monotonic
decreasing function of the GL parameter $\kappa $ is shown. The critical
value $\kappa _{c}=1/\sqrt{2}$ corresponds to $B(0)=1$. The fact that there
exists a monotonic relation between the GL parameter and the magnetic field
at the vortex core center makes it possible to investigate the physical
nature of vortex via $B(0)$ instead of the GL parameter $\kappa $, as we
show below.

In order to study the stability of the vortex, we consider the Gibbs energy
difference between the vortex state under the thermodynamic critical field
and the Meissner state $\Delta G=G_{vortex}(H_{c})-G_{Meissner}$. The same
approach has been used to investigate the stability of vortex in a
two-component superconductor\cite{twocomponent}. With the definition of the
Gibbs energy density $g=f-\mathbf{H}\cdot \mathbf{B}/4\pi $ and equation (%
\ref{energy}), $\Delta G$ can be written in the following form:%
\begin{equation*}
\Delta G=\frac{H_{c}^{2}\lambda ^{2}}{4}\int\nolimits_{0}^{\infty }[\frac{2}{%
\kappa ^{2}}(\frac{d\left\vert \psi \right\vert }{d\rho })^{2}+\frac{1}{%
\left\vert \psi \right\vert ^{2}}(-\frac{dB}{d\rho })^{2}-2\left\vert \psi
\right\vert ^{2}
\end{equation*}%
\begin{equation}
+\left\vert \psi \right\vert ^{4}+(1-B)^{2}]\rho d\rho   \label{gibbs}
\end{equation}

\begin{figure}[tbp]
\begin{center}
\includegraphics[width=3.42in,keepaspectratio]{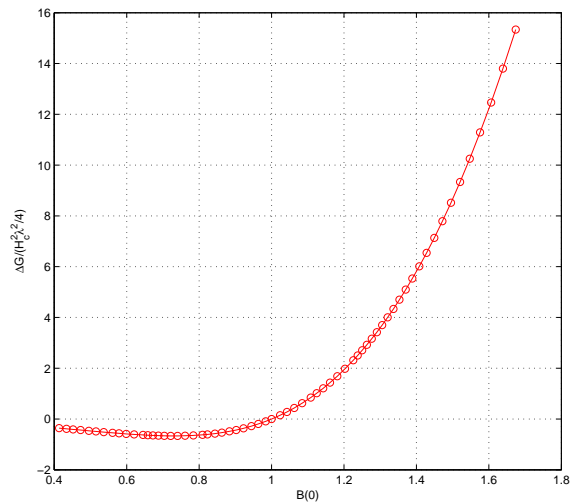}
\end{center}
\caption{(Color online) The Gibbs energy difference $\Delta G$ as function
of the axial magnetic field $B(0)$. $\Delta G=0$ represents the boundary
between type-1/type-2 superconductors, and corresponds to $B(0)=1$.
$\Delta G\rightarrow 0-$ as $B(0)\ll 1$ (see text).}
\label{fig2}
\end{figure}

Note that the Gibbs energy of the vortex state is decreasing with the
increasing field. If $\Delta G>0$, the material always behaves as type-1
superconductor since the Meissner state plays a dominate role. Once the
magnetic flux penetrates into the superconductor in the form of vortex under
certain value of the external field which is smaller than the thermodynamic
critical field $H_{e}<H_{c}$, $\Delta G<0$ is expected. $\Delta G=0$
represents a boundary between the stable/unstable vortex state.

In figure 2, the variation of $\Delta G$ as a function of the magnetic field at
the vortex center is presented. This is a remarkable result. It relates the
local physical quantity, i.e., magnetic field at the center of a vortex,
with the global thermodynamic stability of the vortex state. When $B(0)>1$,
vortex is thermodynamically unstable and $\Delta G>0$, the superconductor is
of type-1; when $B(0)<1$, vortex is thermodynamically stable and $\Delta G<0$%
, the superconductor is of type-2. Zero of $\Delta G$ corresponds to $B(0)=1$.

Asymptotic behavior of $\Delta G$ at $B(0)\ll 1$ can be estimated as
follows. $B(0)$ is a monotonic decreasing function of the GL parameter $%
\kappa $ and $B(0)\ll 1$ corresponds to $\kappa \gg 1$. In this case,
contributions to $\Delta G$ from the vortex core ($\rho <1/\kappa $) and far
range ($\rho >1$) can be ignored. In the region $1/\kappa \ll \rho \ll 1$,
The approximate solution of the magnetic field  is $B(\rho )\approx \sqrt{2}%
/\kappa \ln (1/\rho )$, $\left\vert \psi \right\vert \approx 1$. After some
algebra, we find that, as $B(0)\ll 1$ $(\kappa \gg 1)$, $\Delta G\propto -1/(\sqrt{2}\kappa )\rightarrow 0$.

We then derived the relation between the magnetic field at the vortex core
center and the thermodynamic stability of the vortex. Due to the monotonic
relationship between the core field $B(0)$ and the GL parameter $\kappa $,
one can use $B(0)$ instead of $\kappa $ to investigate the stability of the
vortex. The critical case $B(0)=1$ separate type-1 superconductor in which
there exists no stable vortex $(B(0)>1)$ from type-2 material, in which
there exist stable vortex $(B(0)<1)$.

Although the results of present work are based on the standard GL model
calculations, we believe that, the conclusions have general meaning. The
configurations of the vortices in a charged matter field always bear a
certain resemblance to each other, including a normal vortex core in which
the order parameter is suppressed to zero on the axis of the core, the
restoration of the order parameter and the exclusion of magnetic field far
from the vortex core. Taking into account the fact that the flux carried by
a vortex is quantized, the difference between vortices originates from the
difference of the magnetic field at the vortex core center. And it is
natural that the physical nature of vortex is closely related to the
quantity $B(0)$, the magnetic field at the center of a vortex.

The equivalence of our results and conventional methods which were used to
classify superconductors can be demonstrated as follows. Physical
interpretation of the zero of the surface energy is that the thermodynamic
field is just the critical field at which vortex entry into the
superconductor. The same case can be extended to the vortex geometry, where
we set the external field equals to the thermodynamic critical field.
Besides the emergence of a nontrivial topological charge due to the
variation of the phase of the order parameter around the vortex core, when
the external field equals to the field at the vortex center, $B(0)=1$, $%
\Delta G=0$ can be verified. Moreover, we found that, within the
capabilities of our numerical simulation, no matter what the number of flux
quanta is, field at the vortex core center $B(0)=1$ and $\Delta G=0$ at $\kappa
=1/\sqrt{2}$. This result implies that vortices do not interact in this
regime. All these results proved qualitatively the equivalence of different
types of schemes.

In summary, based on the thermodynamic argument, we have shown that the stability
of vortex in a superconductor, or equally, the classification of the
superconductors according to their magnetic response, has a close relation
with the magnitude of the magnetic field at the vortex center. This finding
provides a new criterion for the classification of the superconductors.
Finally, we note that there exist models in high energy physics which are
mathematically similar to the GL theory for superconductivity (e.g., the
Abelian Higgs model\cite{NO vortex,burzlaff,perivolaropoulos}). Our results can also be generalized to the
investigation of the stability of topological vortex in these models.

\bigskip This work was supported by he Natural Science Foundation of
 Shandong Province (Grant No. ZR2011AQ025) and the National Natural Science
Foundation of China (Grant No. 11104238).


\begin{thebibliography}{99}
\bibitem{code} GL equations (\ref{vortex}) with boundary conditions (\ref%
{infinite}) and (\ref{core}) are boundary value problems of nonlinear
differential equations and can be solved by the collocation method. Note
that we have got the short-range asymptotics of the functions $\left\vert
\psi \right\vert $ and $B$ in Eq. (\ref{core}). It is straightforward to
obtain the derivatives of $\left\vert \psi \right\vert $ and $B$ as
functions of $a$, $B(0)$ from Eq. (\ref{core}) near the core center. The
problem can then be solved with collocation method and $a$, $B(0)$ are
determined. In order to obtain precise solutions for a wider range of
intrinsic parameter, we numerically solved the usual GL Eqs in which the
vector potential $\mathbf{A} $ and the order parameter $\Psi $ are
functions. Then the results are translated to the gauge invariant
quantities $\left\vert \psi \right\vert $ and $B$.
\end{thebibliography}
\end{document}